\documentclass[journal=jctcce, layout=twocolumn]{achemso}
\usepackage{graphicx}  
\usepackage{dcolumn}   
\usepackage{bm}        
\usepackage{amssymb}   
\usepackage{multirow}  

\usepackage{natbib}

\title{The coarse-grained HiRE-RNA model for de novo calculations of  RNA free energy surfaces, folding pathways and complex structure prediction}
\author{Tristan Cragnolini}
\affiliation{Laboratoire de Biochimie Th\'eorique UPR 9080 CNRS, Universit\'e Paris Diderot, Sorbonne, Paris Cit\'e, IBPC
13 rue Pierre et Marie Curie, 75005 Paris, France}
\author{Yoann Laurin}
\affiliation{Laboratoire de Biochimie Th\'eorique UPR 9080 CNRS, Universit\'e Paris Diderot, Sorbonne, Paris Cit\'e, IBPC
13 rue Pierre et Marie Curie, 75005 Paris, France}
\author{Philippe Derreumaux}
\affiliation{Laboratoire de Biochimie Th\'eorique UPR 9080 CNRS, Universit\'e Paris Diderot, Sorbonne, Paris Cit\'e, IBPC
13 rue Pierre et Marie Curie, 75005 Paris, France}
\affiliation{Institut Universitaire de France, Boulevard Saint-Michel, 75005, Paris}
\author{Samuela Pasquali}
\email{samuela.pasquali@ibpc.fr}
\affiliation{Laboratoire de Biochimie Th\'eorique UPR 9080 CNRS, Universit\'e Paris Diderot, Sorbonne, Paris Cit\'e, IBPC
13 rue Pierre et Marie Curie, 75005 Paris, France}

\begin{document}

\begin{abstract}
HiRE-RNA is a simplified, coarse-grained RNA model for the prediction of equilibrium configurations, 
dynamics and thermodynamics. 
Using a reduced set of particles and detailed interactions accounting for base-pairing and stacking we show that non-canonical and multiple base interactions are necessary 
to capture the full physical behavior of complex RNAs.
In this paper we give a full account of the model and we present results on the folding, stability and free energy surfaces of 16 systems with 12 to 76 nucleotides of increasingly complex architectures,
ranging from monomers to dimers, using a total of 850$\mu$s simulation time.
\end{abstract}

\begin{tocentry}
\includegraphics[width=4.75cm]{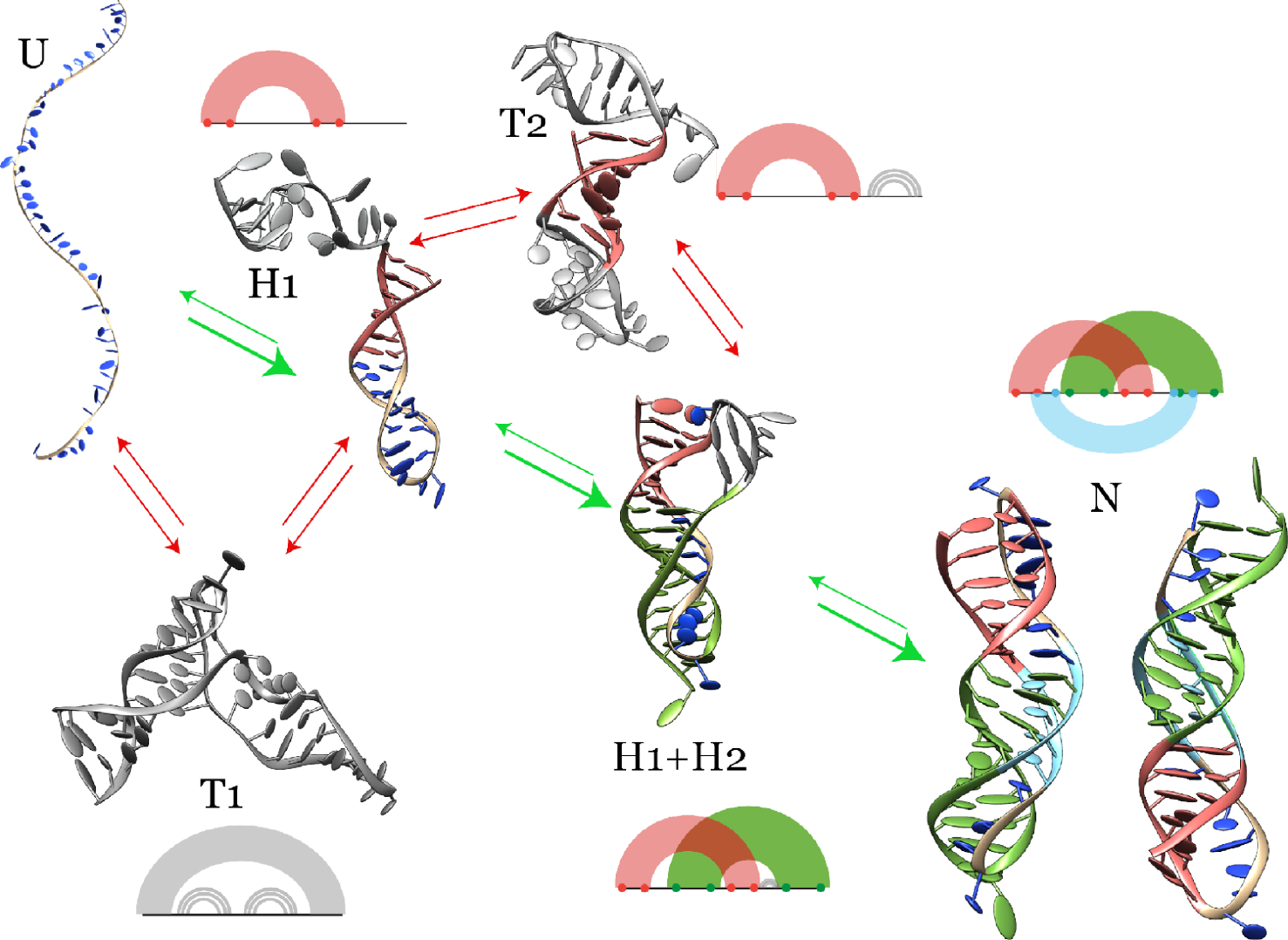}
For Table of Contents Only
\end{tocentry}


\maketitle
RNA molecules are essential cellular machines performing a wide variety of functions for which a specific three-dimensional structure is required.
Aside for their well known roles of genetic information carrier (mRNA) and amino acid recruiter (tRNA), they play a wide range of functions, from regulating gene expression through post-transcriptional 
process (miRNA) and gene silencing (RNAi), to catalytic activities (ribozymes).
Their sizes vary from a few dozen nucleotides for miRNA and RNAi, up to a hundred nucleotides for ribozymes, and to a few thousands for ribosomal RNA constituting the ribosome together with proteins.
In all their diversity these molecule share the common feature of adopting a specific three-dimensional structure to be functional, in the same way proteins must adopt a well defined 3D shape to be able 
to perform their biological activity, posing the question of RNA folding, that is understanding how an RNA linear molecule adopts its characteristic 3D structure.
RNA functionality depends crucially on their equilibrium structures and their dynamical behavior \cite{Holbrook2005, Strobel2008}, 
with distinct active conformations biologically active under different conditions \cite{Scott2007}.

With most of the DNA identified as "non-coding", therefore possibly coding for RNA molecules, being able to determine RNA structures from sequences is essential for our understanding of the cellular machinery.
However, obtaining high-resolution 3D structures through X-ray crystallography and NMR is a 
challenging task as it is proven by the small number of resolved structures in the 
Nucleic Acids Data Bank (NDB) and by the scarcity of structures with substantially different architectures. 
Low-resolution techniques, such as SAXS and Cryo-EM, allow for easier access to the raw data, but require extensive modelling to propose a well-resolved structure.

Computational methods have recently been developed to complement experimental information in the task of predicting 3D structures, following different strategies.
Bioinformatic algorithms based on sequence homology, fragment assembly and secondary structure predictions \cite{Flores2010,Parisien2008,Das2007,Cao2011,Laing2013,Shapiro2007} are 
successful for systems similar to those already present in structural databases. 
They provide a static view of the structure and, sometimes, partial thermodynamical information based on secondary structure predictions, but they are 
not suited for the study of the dynamical and of the global thermodynamical properties of RNA in three dimensions.
These methods base much of their results on the prediction of a secondary structure first, through more or less refined 2D prediction algorithms \cite{Zuker2003a, Xu2014, Rivas1999}.
However, RNA structures are often intricate, giving rise to complex pseudoknots, triple or quadruple base pairings, non-canonical (non Watson-Crick) pairings involving the base's sugar and Hoogsteen 
edges \cite{Leontis2002b}, which are not accounted for in secondary structure prediction methods, developed to address nested structures (tree-like structures) or simple pseudoknots. 

Physical models, considering the interactions of the system's particles in three-dimensions, are better suited to study RNA structure in all of their complexity.
As opposed to secondary structure prediction methods, physical models do not have a specific term for pseudoknot formation. 
A pseudoknot results from the minimization of the free energy and is not encoded as a separate term in the potential energy:
pseudoknots arise from a different organization of base pairings, but the interactions are the same that go into generating a hairpin or any nested structure.

All-atom simulations have successfully folded RNA of 12 nucleotides (nt) \cite{Chen2013, Bowman2008,Zhuang2007}, but are limited to small systems even when adopting an implicit solvent representation \cite{Chakraborty2014}.
Unfolding atomistic simulations have been performed on structures up to about 40 nt \cite{Zhang2011}.
To overcome the limitations imposed by the size of the molecule, and to be able to follow the large scale rearrangements occurring in folding, one can resort to a simplification of the system through coarse-graining.
The challenge of this approach is to design a force-field able to capture all the subtle interactions giving rise to folding, while maintaining a sufficiently simple description of the system for efficient simulation.
Interesting insights on folding mechanisms of structures such as the 49 nt telomerase pseudoknot \cite{Biyun2011} were obtained using Go-like potentials, allowing the formation of native interactions only.
These simulations were performed with a strong bias toward the already known experimental structure, and even though they provide important information on the overall folding process, they are not suited for the prediction of the structure associated to a given sequence, nor for the study of the realm of possible states that the molecule might explore in its life.
Several \emph{ab initio} coarse-grained models, with average interactions between bases and chain connectivity are able to drive the formation of small helical stems and helix, but have limited success for more intricate 3D structures \cite{Ding2008, Hyeon2011, Xia2010, Bernauer2011a, He2013a, Sulc2014}.
Any model aimed at predicting large-scale 3D rearrangements needs to give an accurate description of base-pairing and stacking, including the formation of non-canonical pairs and simultaneous pairings of three or four bases \cite{Chen2009}.
However designing a force field properly accounting for these interactions is a challenging task as shown by the difficulties of otherwise very successful models.
For atomistic simulations Garcia had to re-parameterize the AMBER force field to model correctly three tetraloops of 12 nucleotides \cite{Chen2013}.
The 3D \textit{ab initio} FARFAR procedure, close in spirit to the ROSETTA approach used for proteins, and based on sampling using a coarse-grained model followed by full-atom refinement, cannot reproduce any of the hydrogen bonds or stacking patterns within the UUGC tetraloop \cite{Das2011} and is of limited accuracy for RNAs of 12 to 20 nucleotides in spite of including non-canonical pairings.

HiRE-RNA \cite{Pasquali2010,Cragnolini2013} is an effective theory developed to fold any RNA architecture and study the
 structural dynamics and thermodynamics of RNA molecules.
Through the representation of 6 or 7 beads per nucleotide (figure \ref{fig_HiREv3}), HiRE-RNA v3 force field, with specifically designed energy terms for stacking and base-pairing, 
including non-canonical and multiple pairs, points to the essential physics involved in folding.
Similar to other coarse-grained models, two previous versions of HiRE-RNA, with a less sophisticated force field in the treatment of the bases, were useful to describe small helical stems and duplexes, 
but failed to capture the physics of more complex molecules.
The latest version of the model, HiRE-RNA v3, allows folding of complex systems such as multiple helices and pseudoknots, and provides realistic energy landscapes. 
In contrast to other structure prediction methods such as iFoldRNA \cite{Ding2008}, RNA2D3D \cite{Martinez2008} and MC-Fold/MC-Sym \cite{Parisien2008}, which are limited to the study of one single 
RNA chain, HiRE-RNA can be used to study duplexes, quadruplexes and any multiplexes.

We present here the completely redesigned force field of HiRE-RNA v3 and results on 16 systems spanning from 12 to 76 nucleotides in length, for a total simulation time of  856 microseconds, including several pseudoknots and systems with complex topologies.
For these molecules our model allows to extract thermodynamic information and folding pathways in agreement with experimental data.
Our work highlights the importance of an accurate description of base-pairing in the physical theory, including the plurality of possible base-pairs,
in order to address all the fine structural details playing a key role in the folding process.

\section{Model}
\begin{figure}[t]
  \begin{center}
  \includegraphics[width=0.45\textwidth]{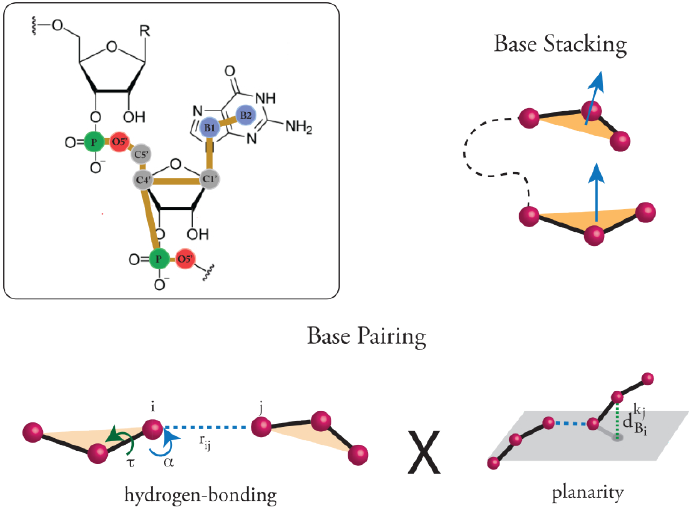}
   \caption{\label{fig_HiREv3} Coarse-grained model and schematic representation of base-pairing interactions, 
   composed of the product of hydrogen bonding and planarity, and stacking, depending on the orientation of vectors normal to the base plane.}
  \end{center}
\end{figure}

HiRE-RNA v3 interaction potential is given by the sum of covalent bond interactions $E_{lc}$, excluded volume $E_{ev}$, electrostatics $E_{el}$,
 stacking $E_{st}$, and base-pairing $E_{bp}$ (Figure \ref{fig_HiREv3}).
Local interactions are described by harmonic terms for bond lengths and angles and sinusoidal terms for dihedrals: 
\begin{eqnarray}
E_{lc} &=&\varepsilon_b\sum_{\mathrm{bond}} (r-r_{eq})^2 + \varepsilon_a\sum_{\mathrm{angle}}  
(\alpha-\alpha_{eq})^2 + \\ \nonumber  && +  \varepsilon_d\sum_{\mathrm{didhedral}} [1 + cos(\tau-\gamma)] ,
\end{eqnarray}
where $r$, $\alpha$ and $\tau$ are the instantaneous bond length, angle and torsion respectively, $r_{eq}$, $\alpha_{eq}$, $\gamma$ are the equilibrium values, and $\varepsilon_b$, $\varepsilon_a$ and$\varepsilon_d$ are the relative strength of the interactions.
Excluded volume is described by a simple decreasing exponential:
\begin{equation}
E_{ev} = \varepsilon_{ev} e^{-\kappa(r-r_v)}.
\end{equation}  
where $r$ is the instantaneous distance between two particles, $r_v$ is a characteristic distance,
and $\kappa$ and $\varepsilon_{ev}$ are the geometric and energetic parameters of the interaction. 

In the absence of explicit ions, electrostatic repulsion of the phosphates and charge screening is represented by a Debye-Huckel potential between P particles:
\begin{equation}
E_{el} = \varepsilon_{el}\frac{q^2}{4\pi \epsilon_0\epsilon_r r}e^{-r/\kappa},
\end{equation}
where q is the elementary charge, $\epsilon_0$ and $\epsilon_r$ are the dielectric constants, $\kappa$ is the Debye length
and $\varepsilon_{el}$ is the adjustable strength of the interaction.

\begin{figure*}[t!]
  \begin{center}
  \includegraphics[width=0.65\textwidth]{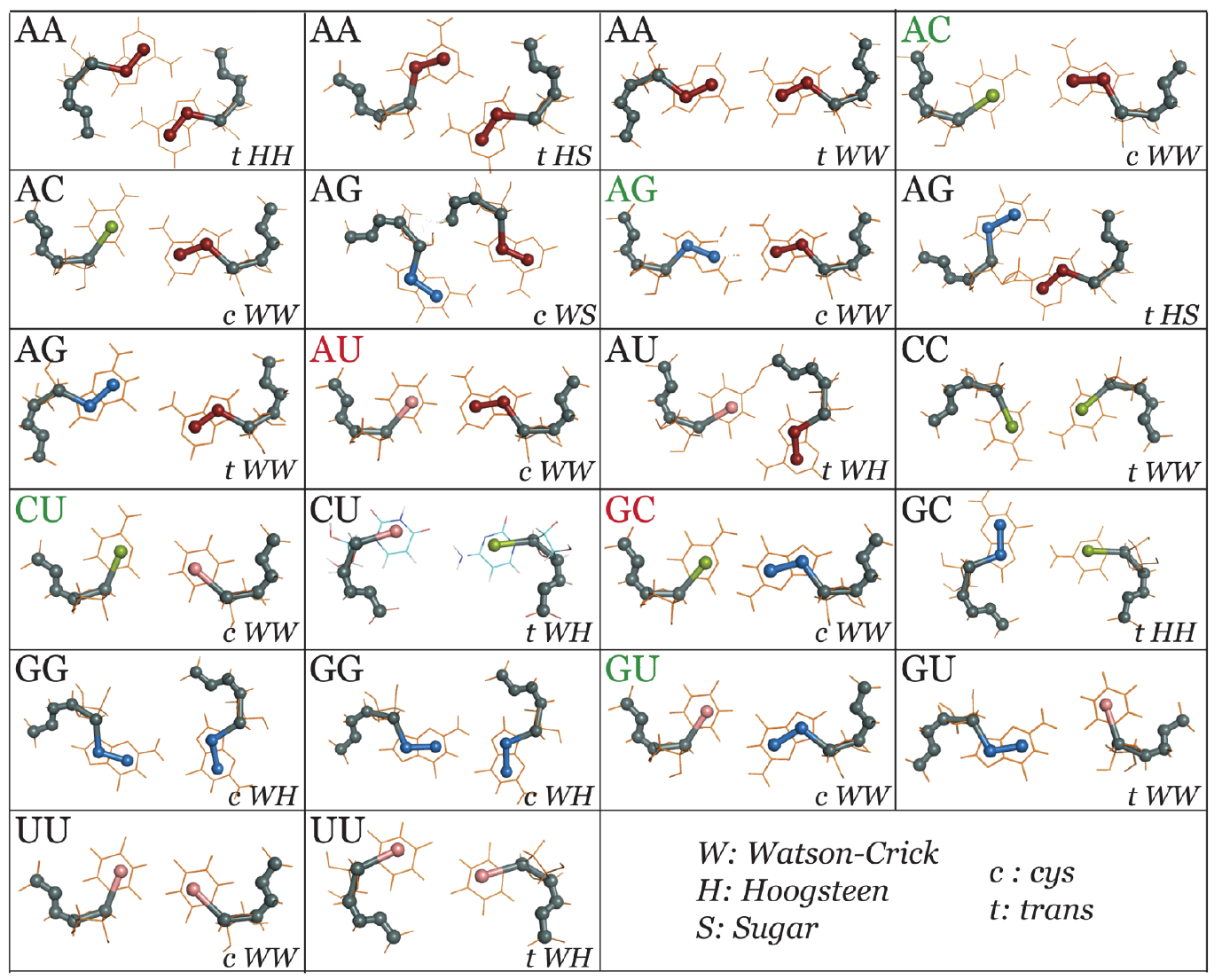}
   \caption{\label{fig_S1} The set of 22 possible base-pairs in HiRE-RNA v3. 
   In red are the canonical WC pairs considered by all CG models, in green are the pairs occurring on the WC sides of the bases, also included in version v1 and v2.}
  \end{center}
\end{figure*}

RNA folding is driven by stacking and hydrogen bonding interactions \cite{Li2008a,Sosnick2003}.
These depend crucially on the relative position and orientation of the bases.
We introduce the concept of base plane, identified through a vector $\vec{n_i}$ normal to the plane defined by the particles B2-B1-CY, 
for bases A and G, and B1-CY-CA for bases C and U. 
Stacking occurs when two particles are close to an equilibrium distance $r_{st}$, when the normal vectors are parallel,
and when the bases are vertically aligned:
\begin{eqnarray}
E_{st} &=& \varepsilon_{st} \; e^{-\frac{(r -r_{st})^2}{\sigma}} \; \left(\vec{n_i}\cdot \vec{n_j}\right)^2  \nonumber \\
&& \left(1-|\vec{n_i} \times \vec{r}\,|^4\right) \left(1-|\vec{n_j} \times \vec{r}\,|^4\right),
\end{eqnarray}
where $\sigma$ and $\varepsilon_{st}$ are adjustable geometric and energetic parameters.
The last two terms allow the bases to form a strong interaction even when they are slightly off-centered to account for the variability in stacking positions.
\\
Base-pairing occurs when two bases are side by side on the same plane, and depends on the relative distance
 and angles of the particles forming the hydrogen bonds. 
We define the base-pairing potential, $E_{bp}$, as 
the product of $E_{bp-pl}$, assuring planarity,  and  of $E_{bp-hb}$ assuring the correct geometry of the interacting particles.
Planarity is imposed by requiring that all particles of one base lie on the plane defined by the other base:
\begin{equation}
E_{bp-pl} = \varepsilon_{pl} \; \left(\sum_{k_j=1}^3 e^{-(d^{k_j}_{B_i}/\delta)^2}\right) \left(\sum_{k_i=1}^3 e^{-(d^{k_i}_{B_j}/\delta)^2}\right),
\end{equation}
where $d^{k_j}_{B_i}$ is the distance of a particle $k$ of base $j$ with respect to the plane of base $i$ and 
$\delta$ and $\varepsilon_{pl}$ are adjustable parameters subject to optimization.
\begin{equation}
E_{bp-hb} = \varepsilon_{hb}\; e^{-(r-\rho)^2/\xi} \;\nu(\alpha_1)\nu(\alpha_2).
\end{equation}

\begin{equation}
\nu(\alpha) = 
\left\{
\begin{array}{ll}
\cos^6(\alpha - \alpha_0), & \mbox{for } -90^\circ \leq \alpha - \alpha_0 \leq 90^\circ ; \\
0, & \mbox{otherwise},
\end{array}
\right .
\end{equation}
where $\rho$ is the equilibrium distance for the pair, and $\alpha_1$ and $\alpha_2$ are the angles formed by the axis of the base and that of the binding particle.
The torsional angle $\tau$ is used to discriminate between interaction minima at $+\alpha$ and $-\alpha$.
To break the symmetry of the cosine function, which would give rise to interaction minima at both $+\alpha$ and $-\alpha$, 
we compute the dihedral angle $\tau$ between the particles defining the base plane of nucleotide $i$ and the interacting bead of base $j$.
If the cosine of the dihedral is negative, $\alpha$ is set to $-\alpha$:
\begin{equation}
\alpha = 
\left\{
\begin{array}{ll}
+\alpha, & \mbox{if } \mathrm{cos}(\tau) > 0 \\
-\alpha, & \mbox{otherwise}.
\end{array}
\right .
\end{equation}

In RNA complex architectures, it is typical to find non-canonical base pairs involving one or more of the three possible sides of the base: Watson-Crick, Hoogsteen, and Sugar.
Bases can than form simultaneous multiple interactions giving rise to triplets and quadruplets.
To represent the wide variety of hydrogen bond patterns
HiRE-RNA v3 includes 22 different base-pairs occurring on all sides of the base (figure \ref{fig_S1}), 
each associated to a specific set of distance, angles, torsions and number of hydrogen bonds formed \cite{Lemieux2002}. 
The choice of 22 interactions is rather arbitrary and can be extended to any number of interactions as long as they are sufficiently distinct in interaction centers.
The pairs included in HiRE-RNA v3 have been chosen based on their abundance in the NDB, making sure to have at least two or three representative for each letter pair.
For some letter pairs we can account for two distinct interaction sites occurring between the same sides at different geometric centers (see \textit{cWW} A$\cdot$C pairs).
For any given letter pair we consider all possible base-pairs from our list by adding over all possible $E_{hb}$ terms.
Because of the narrow distance dependence of $E_{hb}$ and of the excluded volumes of the beads, effectively, there can only be three interaction centers simultaneously 
present around a base, one on each side.

\begin{figure}[h]
  \begin{center}
   \includegraphics[width=0.45\textwidth]{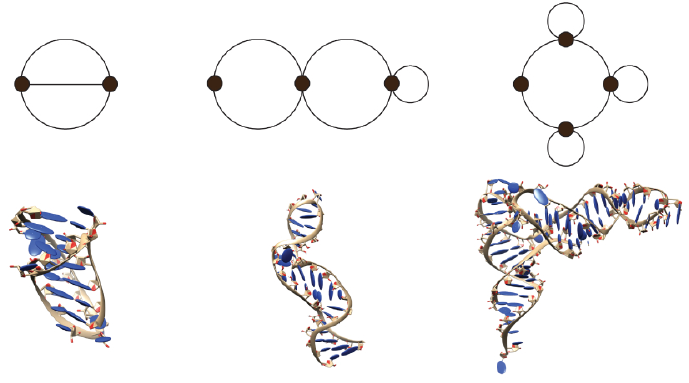}	
   \caption{\label{fig_dual} Dual graph representation of structures with different topologies: simple pesudoknot (left), helices and hairpin (center), 4-way junction (right).
   The 3D structure of an example of each topology is given under the dual graph (PDB ID: 2A43, 1A9L, 1L9V).}
  \end{center}
\end{figure}

The model has geometric parameters whose values have been determined from distributions extracted from 200 NDB structures including molecules of varying sizes and 
topologies, and overall energetic parameters, representing the relative weights of the different interaction terms, which are subject to an optimization procedure.
We used the concepts of RNA graphs to build a structure database rich in different topologies  
since this descriptor captures well the different overall organization of the molecule's structure \cite{Gan2003,Pasquali2005}.
Through ``dual graphs'' (figure \ref{fig_dual}), the RAG database \cite{Izzo2011} enumerates graphs corresponding to possible RNA topologies, including pseudoknots 
of arbitrary complexity, giving a link to the corresponding PDB structure, when this exists.
From this database, we have chosen one an equal number of representative structures for each populated topology, 
to form a training set of 20 RNAs used to optimize parameters through a genetic algorithm \cite{Maupetit2007a}.
For each native structure we first generate an ensemble of decoys varying in RMSD and number of native contact to cover the four possible scenarios of low-RMSD/high native BP, low-RMSD/low native BP, 
high-RMSD/high native BP and high-RMSD/low native BP.
The algorithm then selects parameters maximizing the energy difference between native configurations and decoys.
Parameters obtained with this procedure were then tested through long MD simulations on systems of various sizes and showed a significant improvement over the previous parameters 
calibrated by hand both on folding predictions and on stability tests.
The new parameters were validated on several molecules not included in the training set, 
in their ability to fold small hairpins and to give the correct melting temperatures for duplexes.

\section{Results}
\begin{table*} [tb]
\centering
\begin{tabular}{ c c c c c c c}
 {\bf PDB} & {\bf Topo} & {\bf Nb} &  {\bf Method} & {\bf RMSD} & {\bf BP} & {\bf Time}\\
  &  & nt & & \AA & tot (obs/nat) & $\mu s$\\
\hline
1F9L &         Hp        &  22    &     ST              &  3.2         &    10 (9/9)     &   3      \\
1L2X* &         Pk        &  27    &    REMD            & 5.8          &    11 (7/8)     &   134     \\
1N8X &         Hp/Bl     &  36    &    REMD             &  3.8         &    15 (11/15)     &   26    \\
1RNG &         Hp        &  12    &     ST              &  2.7         &     5 (5/5)       &   3       \\
1ZIH &         Hp        &  12    &     ST              &  1.9         &     5 (5/5)       &   3       \\
1F7Y &         Hp        &  12     &     ST             & 2.0         &    5 (5/5)       &  3      \\
1Y26* &         C         &  71    &    REMD            &  8.1         &    32 (15/29)     &  96     \\
2G1W &         Pk        &  22    &    REMD             &  4.3         &    9 (7/7)       &   153    \\
2G1W &         Pk        &  22    &     ST              &  4.4         &    8 (7/7)       &   3      \\
2K96 &         Pk-3Hx    &  47    &    REMD             &  4.3         &    23 (17/22)     &  120   \\
480D &         Hp        &  27    &    REMD             &  5.5         &    12 (9/9)      &  180    \\
405D &         Dp        &  2x16  &    REMD             &  3.6         &    16 (12/16)     &  64   \\
433D &         Dp        &  2x14  &    REMD             &  4.0         &    14 (14/14)     &  64   \\
\hline
3L0U* &         tRNA      &  73    & 	MD 		&  $\sim$ 8    & 32 (23/30)     &   0.5   \\
6TNA* &         tRNA      &  76    & 	MD 		&  $\sim$ 8    & 31 (23/29)     &   0.5   \\
1KF1* &        G-quad     &  22    & 	MD  		&  $\sim$ 4    & 14 (12/12)       &   3   \\
\end{tabular}
\caption{\label{systems_table} Summary of the systems studied. For each system we give PDB entry, topology
 (Hp: hairpin, Pk: pseudoknot, Bl: bulge, C: complex, Hx: helix, Dp: duplex, G-quad: G-quadruplex),
 number of nucleotides, simulation method, RMSD of the center of the most populated cluster at 300 K using REMD/ST or of the structure after several hundreds nanoseconds,  total
number of base pairs and number of native pairs found in the predicted and experimental structures, total simulation time.  
The top simulations started from a fully extended structure, while the last three simulations give the results of MD stability 
starting from the NMR structures.
Structures with a * are experimentally determined in conditions that we cannot fully include in our simulations: 1Y26 contains the adenine ligand; 
3LOU, 1L2X and 1KF1 contain structural ions, 6TNA contains modified bases.
}
\end{table*}

Table~\ref{systems_table} reports the folding and stability results on 16 topologically different RNA molecules of 12 to 76 nts.
For each system we performed either Molecular Dynamic (MD) simulations at 300 K, Replica Exchange Molecular Dynamics (REMD),
 64 replicas
with temperatures from 200 K to 400 K and  500-1200 ns per replica, or Simulated Tempering (ST)  with temperatures from 300 K to 500 K,
all  with an integration time-step of 4 fs and a Langevin thermostat \cite{Sterpone2014}.
For analysis, we monitored native base pairs and root mean square deviation (RMSD) from the native structure using all particles.

\begin{figure}[t!]
  \begin{center}
   \includegraphics[width=0.45\textwidth]{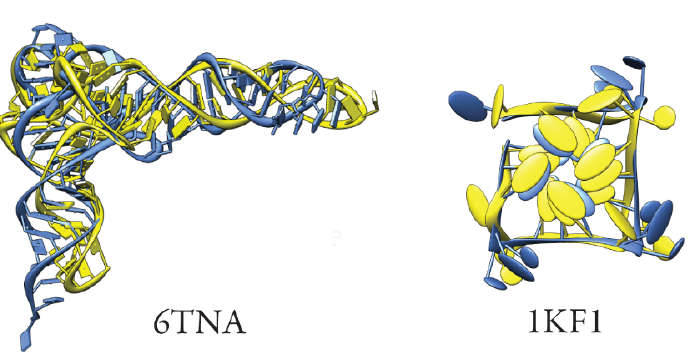}	
   \caption{\label{fig_stability} Superposition of the native structure (yellow) and a structure from the MD stability simulation (blue) after several hundred nanoseconds of simulation time.}
  \end{center}
\end{figure}

With HiRE-RNA v3 we have been able to fold 13 structures from fully extended configurations within a few RMSD from the experimental structure and reproducing most of the native base-pairing network.
To test the extent of the validity of our force field, we also performed long MD stability simulations on three systems for which complete folding from fully extended states is at the moment out of reach because of the presence of structural ions or of modified bases which we can't account for in our model at this stage.
We have analyzed G-quadruplexes, found on telomeres and of importance for cancer regulations, bound to several intercalating ions, and  
two tRNAs with both modified bases and structural ions.
Despite the absence of specific interactions for structural ions and modified bases, the three systems do not depart significantly from their NMR structures over
 several hundred nanoseconds,
the quadruplex remaining at 4.0\AA~ for over 3$\mu s$, and the two tRNA of 73 and 76 nts remaining at 8.0\AA, with the native architecture
preserved and most native base pairs formed (figure \ref{fig_stability}).

To compare the accuracy of HiRE-RNA v3 to atomistic approaches, we first examined three tetraloops studied by the Garcia's group in 2013 using
extensive all atom REMD simulations in explicit solvent \cite{Chen2013}. 
To achieve high-accuracy folded structure,  Garcia et al. had to re-parametrize AMBER forcefield.
By using REMD and ST simulations, we fold the tetraloops with the same RMSD accuracy (1RNG: Garcia 3.1 \AA~ vs. 2.7\AA~ here,
1ZIH: 1.3\AA~ Garcia vs. 1.9\AA~ here, 1F7Y: Garcia 0.8 vs. 2.0 A here),
but with a much smaller computational cost (only 60 CPU hours per run, for 3 microseconds of simulated time).
Importantly for statistics, each ST simulation displays many folding/unfolding events.
The importance of understanding tetraloop formation is reported in a recent article by Wales' group \cite{Chakraborty2014} where the discrete path sampling method was used to determine the folding mechanisms and kinetics of three RNA tetraloops.

\subsection{Small pseudoknots}
Predicting small pseudoknots is particularly challenging because these molecules fold back on themselves forming tight bends,
stabilized extensively by stacking interactions beside base-pairings.
While our v1 and v2 models lacked a detailed stacking term and could not predict such structures, HiRE-RNA v3 can fold
the 22~nt 2G1W \cite{Nonin-Lecomte2006} and 28~nt 1L2X \cite{Egli2002} pseudoknots with the experimental fold as the most stable structure
 at 300 K.

\begin{figure*}[t!]
  \begin{center}
\includegraphics[width=0.8\textwidth]{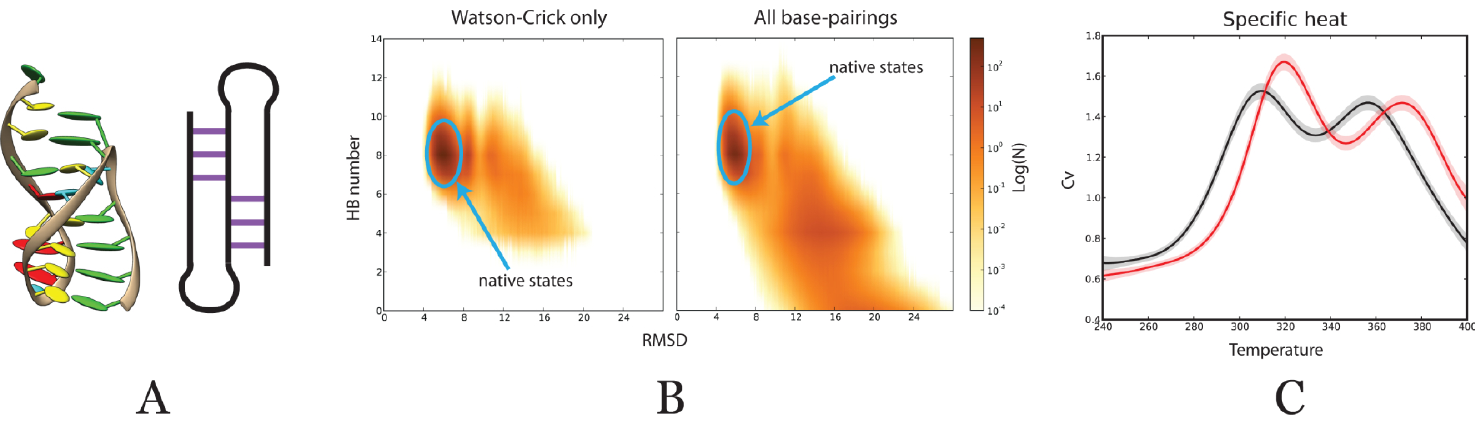}
   \caption{\label{fig_2G1W} A: Predicted 3D structure of 2G1W and 2D representation of the pseudoknot. 
B: PMF (RMSD vs. bp) at 300K of 2G1W considering only WC base-pairs (left) and the full set of possible base pairs (right). The pseudoknot structure (blue insert) is the
 the most populated state in both cases, but when considering the full set of base-pairs a plurality of partially folded/misfolded states is also present.
C: Specific heat curves for 2G1W with WC-only base pairs (red) and with the full set of non-canonical pairs (black).
 The lowest peak corresponds to the transition to the native state, while the highest peak corresponds the transition from a variety of partially folded states to the free chain.  
 Both systems exhibit similar behavior, but melting temperatures are lower when non-canonical pairs are considered. 
 Cv curves are computed using the MBAR algorithm \cite{Chodera2007}; shaded area represent error bars.}
  \end{center}
  \end{figure*}

To study the impact of non-canonical base pairing on the system's behavior, for 2G1W we also performed simulations considering only base pairs on the WC side of the base (red and green base-pairs in figure \ref{fig_S1}).
While the molecule is still able to reach the native state, the absence of non-canonical pairs involving
Hoogsteen and Sugar edges alters the energy landscape (figure~\ref{fig_2G1W} B).

With the full set of base pairs more states are populated as partial intermediates creating a continuous path from misfolded to folded state.
Physically, this is an important difference as it allows the molecule to more easily interconvert between different states compared to when only WC base pairs are considered.
This aspect is crucial for RNAs that are known to adopt alternative architectures at various stages of their biological activity.
The importance of including non-canonical pairs manifests also in thermodynamics.
From our REMD simulations, the estimated melting temperature of the model with all possible pairs is lower than that
 of the model including only WC base pairs (Figure~\ref{fig_2G1W} C). 
 This is indeed expected as the presence of non-canonical pairs opens to the possibility of new folding pathways, absent when only canonical base-pairs are considered, and
 renders the ground state more entropically accessible, lowering the melting temperature.
By shifting the temperature of one of the two curves to superpose the melting peaks we can observe that the low and high temperature behavior of the two systems is the same, but that the lowest peak for the WC-only system is narrower and taller than the corresponding peak for the full base-pairing set, indicating a stiffer transition when non-canonical parings are turned off. 
The computed average energies at all temperatures are the same within error bars, while the fluctuation behavior of the two system is different, with the system including non-canonical pairs subject to larger fluctuations than the system with WC only interactions.

Despite the fact that our model is not explicitly optimized on thermodynamic data, the melting temperatures of the pseudoknot of 2G1W (lower temperature peak)
estimated at 320K and 310K using WC-only and all pairs respectively, 
are in good semi-quantitative agreement with the experimental value of 329K at 50 mM NaCl \cite{Nonin-Lecomte2006}.
Given our model does not explicitly include ions nor water, we don't expect at this stage to have a precise quantitative correspondence between computed and experimental melting 
temperatures, but to be able to predict general features.

\begin{figure}[t!]
  \begin{center}
   \includegraphics[width=0.45\textwidth]{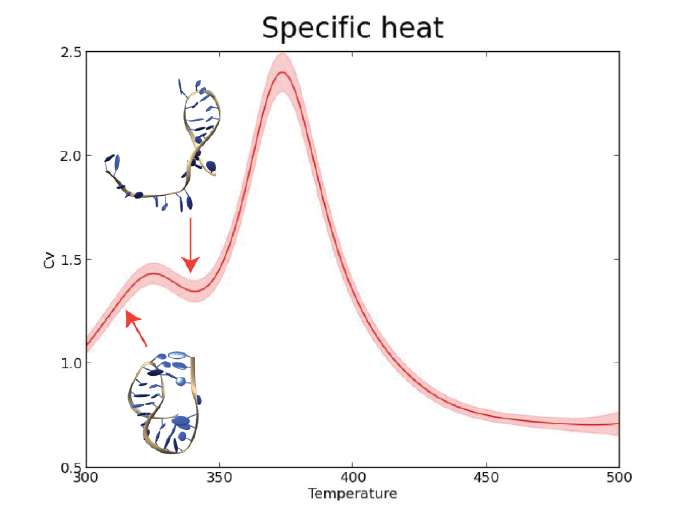}	
   \caption{\label{fig_1L2X} Specific heat curve of 1L2X exhibiting two peaks corresponding to melting of the two stems composing the pseudoknot. The shorter stem melts at lower temperature than the longer stem. }
  \end{center}
\end{figure}

The presence of two peaks in the specific heat curve is in good agreement with the experimental observations on the MMTV pseudoknot \cite{Theimer2000}, a 32-nt RNA with
 the same topology as 2G1W, and seems to be a common feature of pseudoknot folding, where melting of two separate stems is involved.
The specific heat curve of 1L2X also displays two peaks (figure \ref{fig_1L2X}). 
1L2X is a pseudo-knot composed of a longer stem (8 base pairs) and a shorter stem (3 base pairs). 
The lower-temperature peak at 330K ($T_{m_1}$) corresponds to melting of the shorter stem and the higher-temperature peak at 370K ($T_{m_2}$) corresponds to complete unfolding. 
As expected, $T_{m_1}$ is higher than the corresponding melting temperature of 2G1W (320 K) given the higher number of base pairs breaking in the melting transition in 1L2X over 2G1W. 
The calculated difference between $T_{m_1}$ and $T_{m2}$ of 40 K is 
consistent with experimental observations for similar pseudoknots for which differences $T_{m_2}-T_{m_1}$ measured by UV absorbance was reported to be of 20 K, 24 K and 35 K (MMTV pseudoknot, 
wild-type T4 pseudoknot and the C8U BWYV pseudoknot respectively \cite{Theimer2000, Theimer1998}).

\subsection{Triple helix pseudoknot}
\begin{figure*}[t!]
  \begin{center}
   \includegraphics[width=0.6\textwidth]{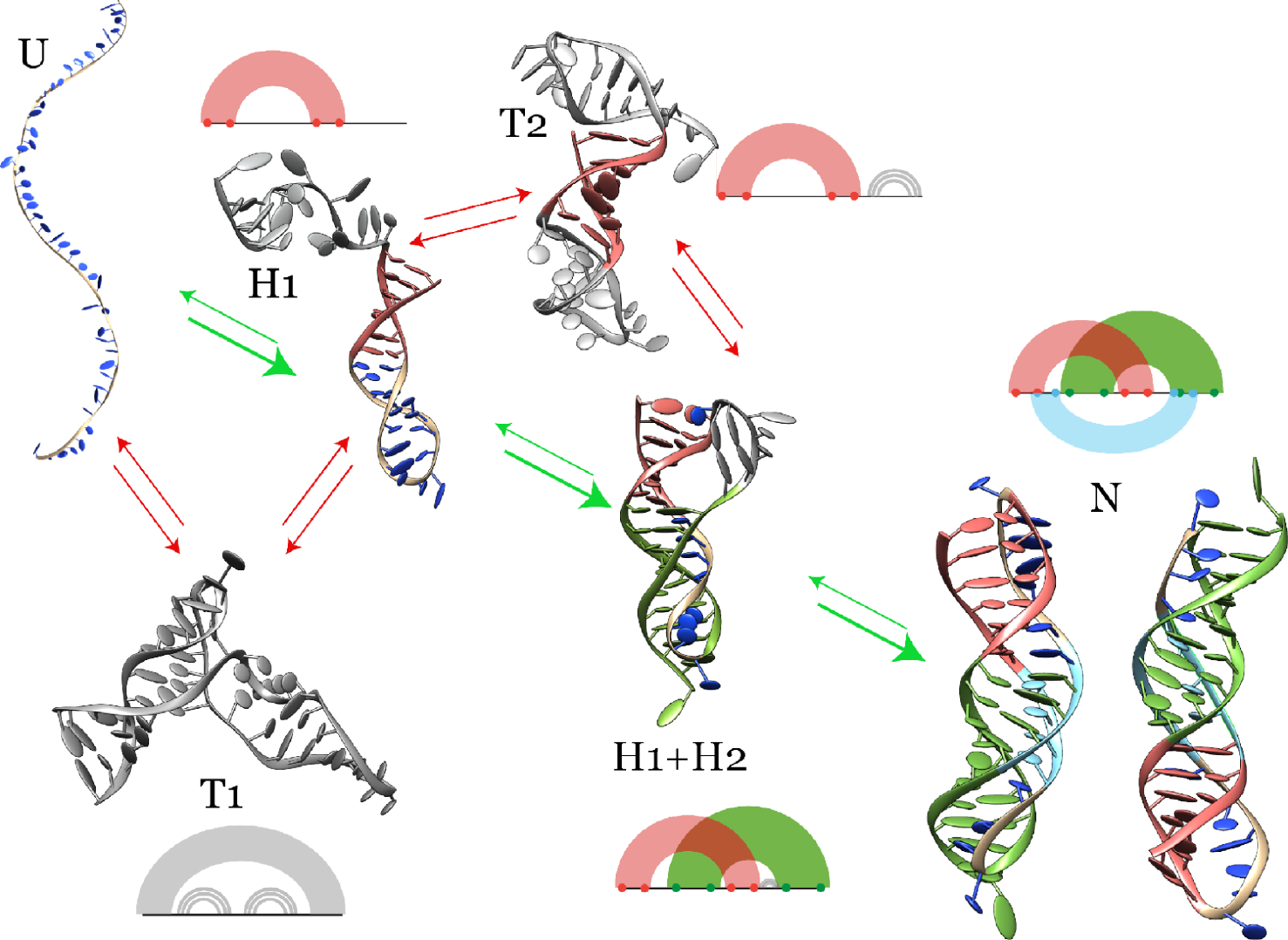}	
   \caption{\label{fig_2K96} Folding pathway of the triple helix 2K96 extracted from REMD simulations. 
Green arrows indicate the reversible transitions between intermediates folds leading to the native structure (N), red arrows indicate transitions leading to misfolded states (T1 and T2). Next to each structure we give the schematic representation of base pairing with arc diagrams color coded according to the formation of H1 (red), H2 (green) and H3 (blue). Structures in gray correspond to non-native secondary elements.}
  \end{center}
\end{figure*}

We then studied larger and more complex systems.
Starting from fully extended conformations, REMD simulations were able to fold the triple helix of the pseudoknot of the human telomerase (PDB id: 2K96) of 47~nt \cite{Kim2008}, 
for which the native structure is characterized by a 6 base-pairs WC helix and an A-rich dangling strand inserting into the WC helix groove and forming several stacked triplets.
Overall, 22 base pairs, canonical and non-canonical, stabilize the native structure.
\begin{figure}[t!]
  \begin{center}
   \includegraphics[width=0.45\textwidth]{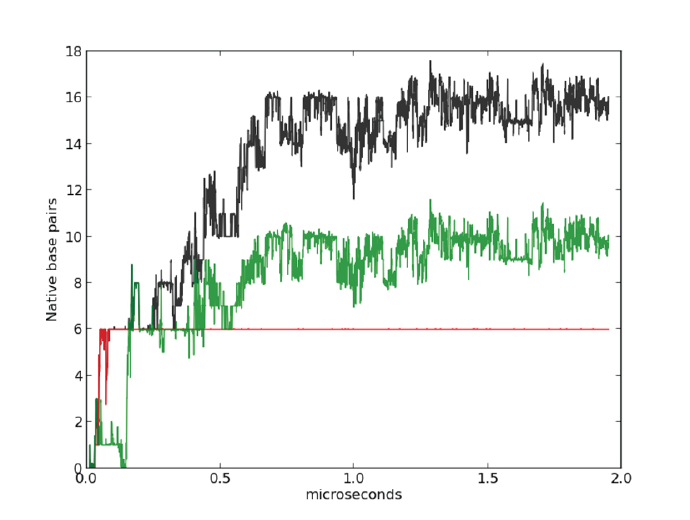}	
   \caption{\label{fig_2K96_HB} Formation of native contacts for the triple helix folding. Contacts of the WC helix (red) form first while contacts stabilizing the triple 
   helix (green) form on a much longer time scale. The full set of native contacts, including those of the triple helix, is shown in black.}
  \end{center}
\end{figure}
In the simulation we can distinguish a short phase to form the WC helix, and a longer phase in which the other contacts form, generating the full triple helix (figure \ref{fig_2K96_HB}). 
After 1.2$ \mu$s REMD time, a structure was reached with an RMSD of 4.3\AA, stabilized by 17 native base pairs (Figure ~\ref{fig_2K96}: N). 
To our knowledge this is the first time anyone has folded an RNA of such complexity solely from the sequence, an achievement possible only if the relevant physics is correctly taken into account by the model.

Although individual ST and REMD individual trajectories rapidly swap temperatures and the observed pathways may not be identical to the pathways observed at constant temperature, significant insights can be obtained from the many unfolding/folding events we can observe in enhanced sampling simulations.
We have therefore further analyzed the folding path of the triple helix 2K96 (figure~\ref{fig_2K96}). 
In the NDB structure of 2K96, the paired regions are the following: 1-6 paired with 24-29 (H1), 15-23 paired with 37-47 (H2) and 6-10 paired with 36-40 (H3). 
Notice that there is an overlap of two paired regions with bases 37, 38, 39 and 40 forming triple contacts. 
In REMD simulations we observe three separate folding steps corresponding 
to the successive formation of each one of the stems (H1-red first, H2-green second, H3-light blue third), with different time scales involved.
The system can remain trapped in misfolded states with structures exhibiting base-pairings different from native (T1 and T2).
Our folding path (H1 $\to$ H1+H2 $\to$ N) is in agreement with experimental studies on 2K96 \cite{Theimer2005}.
Based on UV melting curves, it was proposed that the three melting transitions of increasing temperature correlate with the loss of tertiary structure,
followed by melting of the AU rich stem 2, and eventually loss of the structure in G-C rich stem 1. 
Our results are also in agreement with results obtained by Langevin simulations and a coarse-grained model with Go-like properties (TIS), where the formation of the stems 
and the assembly mechanisms of RNA pseudoknots are determined by the stabilities of constituent secondary structures:
if the secondary structural elements have comparable stability, then there are multiple routes to the native state, 
otherwise there is one dominant path \cite{Cho2009}. 
In the case of 2K96, H1 consists of G$\cdot$C WC interactions, more stabilizing than the A$\cdot$U interactions of H2. 
Our predictions on the folding pathway extend those of the Go-like model study which did not take into account the formation of non-canonical and triple pairings.

\begin{figure*}[t!]
  \begin{center}
   \includegraphics[width=0.8\textwidth]{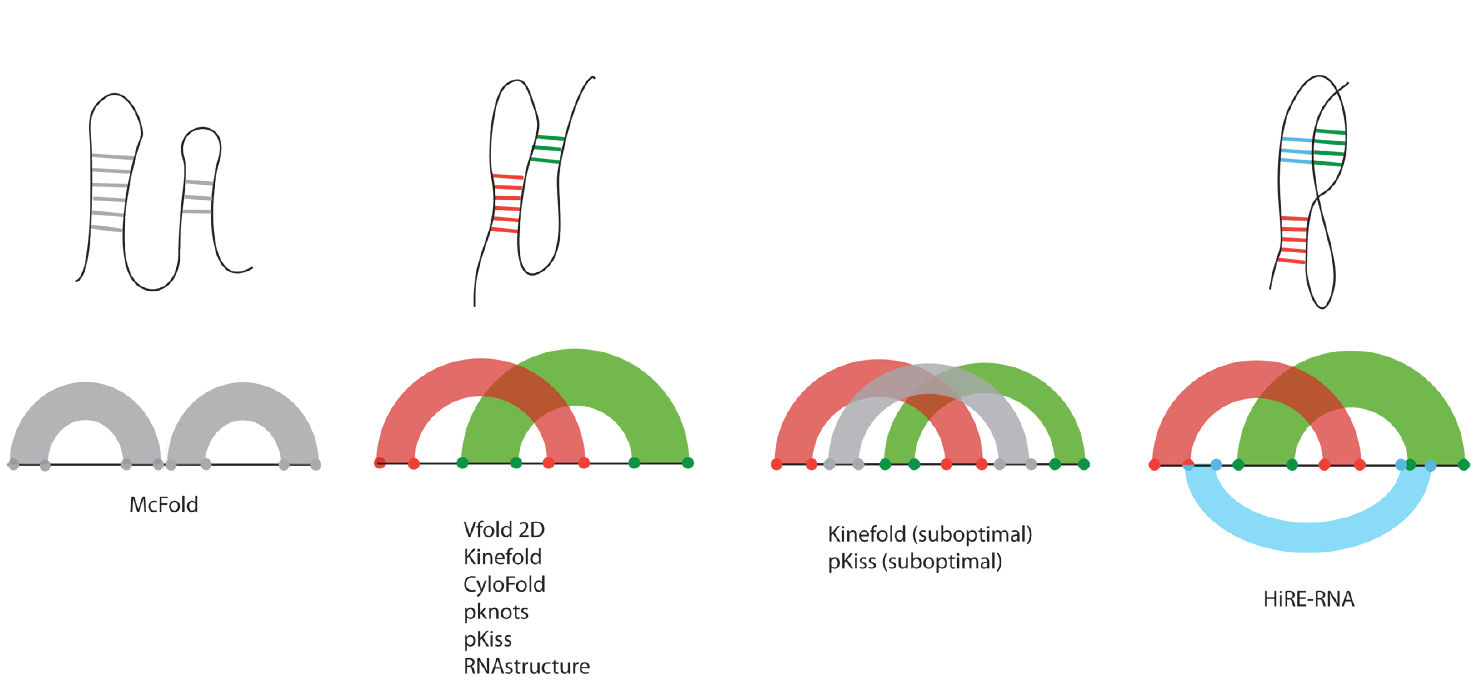}	
   \caption{\label{fig_2K96_pk} Results of secondary structure prediction algorithms for the triple helix pseudoknot.
Under each topology, represented as arc graphs and sketched as secondary structure elements, we list the names of the algorithm that propose that result as optimal or suboptimal prediction.}
  \end{center}
\end{figure*}
Given the challenges of folding a triple helix, for 2K96 we performed an extensive comparison of our results with seven secondary structure prediction methods available on-line all allowing formation of pseudoknots.
We tested the widely used MCFold \cite{Parisien2008}, Kinefold \cite{Xayaphoummine2003}, RNAstructure \cite{Bellaousov2013}, Vfold2D with the Turner’s 
parameters or MFOLD2.3 \cite{Xu2014a}, considered to be the best performing algorithms thus far,  pknot \cite{Rivas1999}, pKiss \cite{Theis2010} and CyloFold \cite{Bindewald2010}.
As it can be seen on figure \ref{fig_2K96_pk} none of these methods predicts the triple helix. 
MCFold predicts two disjoint hairpins and all other methods, considering the optimal solution or suboptimal solutions, predict the simple pseudoknot (pseudoknot H), when explicitly 
instructed to look for a pseudoknot. 
In principle, Vfold2D can predict structures including base triplets, but it does not give the correct result for 2K96 despite the fact that it was shown to be able to produce the correct 
2D structure, including triple contacts, for a similar system \cite{Cao2010}.

\subsection{Large systems}
\begin{figure}[t!]
  \begin{center}
   \includegraphics[width=0.40\textwidth]{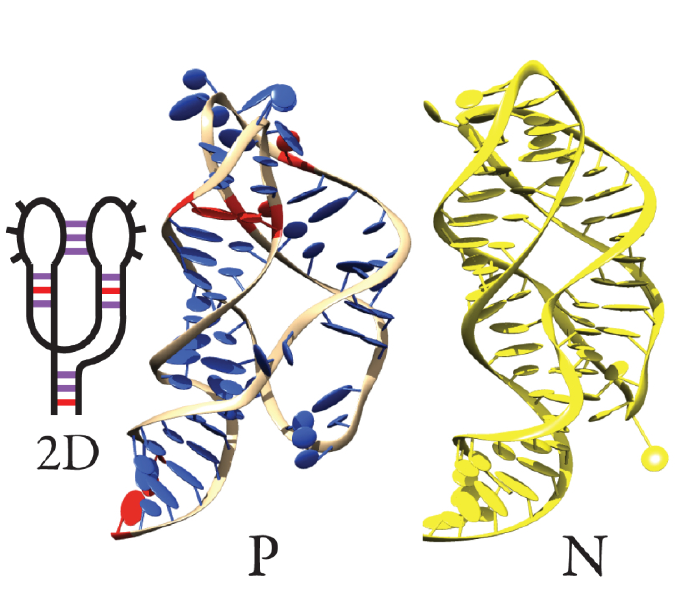}	
   \caption{\label{fig_1Y26} 
1Y26 constrained prediction at 7.1\AA~(P) and native structure in yellow (N). The three local constraints are shown is red in 2D.}
  \end{center}
\end{figure}

Despite the already substantial reduction in degrees of freedom of our theory, folding large structures remains challenging because of the long times needed for accurate sampling.
Folding times can be reduced by adding partial experimental information such as a few base-pairs from NMR or SHAPE \cite{Weeks2010}.
This is the strategy we adopted for the riboswitch 1Y26 of 76~nt starting from a fully extended state 
\cite{Serganov2004}.
In its NMR state with an adenine ligand, 1Y26 adopts a Y shape with the two upper stems binding through kissing loops.
Imposing three base pairs restraints (one WC pair on each helix) taken from VFold2D predictions,
both simple MD (at 300K) and REMD recover the overall organization of the kissing loops, with an RMSD of 7-8\AA (Figure~\ref{fig_1Y26}).
The major discrepancy between our predicted and the NMR structures is at the junction where the adenine ligand, absent in our 
simulations, should sit.
Our results for 1Y26 are comparable in quality to a prior prediction obtained with the automated 3DRNA program, 
based on secondary structure reconstruction \cite{Zhao2012}.
In this program, the smallest secondary elements (SSEs) are assembled into hairpins, 
hairpin loop, internal loop, bulge loop, pseudoknot loop and junction and then  
a network representation of the secondary structure is used to 
describe the locations and connectivity of the SSEs.
Using the whole secondary structure extracted from the experimental 
structure, i.e. much more experimental information than the 3 base-pairing constraints we impose with HiRE-RNA, 
the 3DRNA recovers the structure of 1Y26 with a RMSD of 6.7 \AA.

\section{Discussion}
We have presented here an effective theory for RNA folding, based on a detailed and new description of the two main driving forces, stacking and base-pairing interactions,
able to fold structurally diverse RNAs to their native states when coupled to REMD or ST.
As the need to properly consider base pairing and stacking is clear to anyone working on RNA, how to actually 
define functions to describe these interactions in a simplified representation, allowing to study large-scale rearrangements, is less than straightforward. 
The force field we define is nothing close to any other existing coarse-grained model.
HiRE-RNA v3 force field shares with the previous v2 version the particle description and the functional forms of local interactions,
but it is completely redesigned for all other
terms including new analytic energy functions for base pairing and base stacking 
that allow accurate prediction of equilibrium configurations and thermodynamics of 13 systems with 12 to 76 nucleotides starting from fully extended states. 
The model makes use of a large number of parameters which have now been optimized following a rigorous procedure, a particularly complex task when the possible interactions considered go beyond simple Watson-Crick pairing and address a multitude of possible states including non-canonical and multiple base-pairs \cite{Bottaro2014}.

In spite of its simplicity, similar to Garcia's atomistic simulations, our model predicts the high resolution structure of tetraloops, which are not recovered by FARFAR/FARNA by fragment reconstructions including non-canonical pairs.
On larger structures, the prediction capabilities of HiRE-RNA are comparable to those of the most advanced methods currently used for RNA structure prediction.
However, while we obtain most of our results with no or minimal external information other than the sequence, these methods typically require a substantial input of experimental evidence.
As it was shown in the 2012 RNA-puzzle competition \cite{Cruz2012}, the best performing prediction algorithms so far are those based on fragment reconstruction, giving access to 3D structures, 
but not giving information on other aspects such as the folding pathways and energy landscapes.
In the competition, a riboswitch of 86 nt was predicted by eight research groups with RMSDs ranging from
7.2 to 23.0 \AA~. 
The model of lowest RMSD (7.2\AA) used a multi-scale approach based on 2D structure prediction methods and self-assembly of fragments selected from the NDB \cite{Cao2011}. 
Coarse-grained \textit{ab initio} methods performed much poorly with the best RMSD at 11.5\AA, even when relying heavily on experimental constraints \cite{Ding2008}.
More recently, Xia with his coarse-grained model obtained a structure  at 7.6\AA, using secondary structure prediction tools and 14 constraints \cite{Xia2013}.
As shown by the folding of the triple helix, in the analysis of folding pathways, our model goes beyond what was done through Go-like models \cite{Biyun2011, Feng2011}, given we do not introduce any bias toward the native structure and we account for multiple pairings.

A key feature of our model, distinguishing it from most other methods, is the possibility of forming non-canonical and multiple base-pairs.
Our results show the importance of considering non-canonical base pairs.
Indeed they are essential to fold complex molecules and should not be neglected even for RNAs whose experimental structures contain only canonical pairs
as they have a significant impact on the free energy profile of the system, on its thermodynamics, and possibly on folding pathways.
The presence of non-canonical pairs gives rise to an increase of transition states that can favor interconversion between different configurations, 
a behavior that is observed for many biologically active RNAs \cite{Fuertig2008} and that has been hypothesized for the switch within pseudoknot domain of human telomerase RNA between the pseudoknot 
and hairpin conformations \cite{Theimer2005}.
We are currently investigating these aspects in more detail through disconnectivity graphs \cite{Wales2002} on several pseudoknots. 

As the current status of HiRE-RNA v3 already represents a significant advancement in the study of RNA molecules both in terms of prediction capabilities and in the possibility of addressing
questions concerning dynamical and thermodynamical behaviour, we are extending our work to include other interaction terms 
such as the presence of ions and ligands, in order to be closer to experimental conditions.
Combined with the coarse-grained OPEP force field \cite{Sterpone2014}, HiRE-RNA should help understand the interplay between proteins and nucleic acids.

\begin{acknowledgement}
This work was supported in part by the "Initiative d'Excellence" program from the French State (Grant "DYNAMO", ANR-11-LABX-0011-01)" 
and IUF.
\end{acknowledgement}

\bibliography{publis_v3}

\newpage
\end{document}